\documentclass[a4paper,11pt]{article}
\pdfoutput=1
\usepackage{amsmath}
\usepackage{amssymb}
\usepackage{color}
\usepackage{cite}
\usepackage{hyperref}
\usepackage{graphicx}
 \usepackage{leftidx}
 \usepackage{subfig}
\usepackage{showlabels}

\numberwithin{equation}{section}

\def \be {\begin{equation}}
\def \ee {\end{equation}}
\def \ba {\begin{array}}
\def \ea {\end{array}}
\def \bea{\begin{eqnarray}}
\def \eea{\end{eqnarray}}

\def \D {\Delta}

\begin{document}

\title{\textbf{A further support for the statement that the action of the HRT-area generates a kink transformation in the pure AdS$_3$ gravity}}
\author{
Mai Qi$^{1}$\footnote{maiqi@ucsb.edu}\,
Xiao-Shuai Wang$^{2,3}$\footnote{wangxiaoshuai@itp.ac.cn}\,
Jie-qiang Wu$^{2,3}$\footnote{jieqiangwu@itp.ac.cn}
}
\date{}

\maketitle

\begin{center}
{\it
$^{1}$College of Creative Studies, University of California, Santa Barbara, CA, 93106, U.S.A \\
$^{2}$Institute of Theoretical Physics, Chinese Academy of Sciences, Beijing 100190, China \\
$^{3}$School of Physical Sciences, University of Chinese Academy of Sciences, \\ Beijing 100049, China
}
\vspace{10mm}
\end{center}

\begin{abstract}

It is stated that the action of the Hubeny-Rangamani-Takayanagi (HRT) area, which is viewed as an observable, generates a kink transformation in gravity. In this paper, we provide a further support for the kink transformation statement by performing a relevant computation.

Specifically, by acting the kink transformation to the boundary stress tensor, we compute the bracket between the HRT area with the boundary stress tensor. Here we represent the boundary stress tensor in terms of initial data on the Cauchy surface under holographic renormalization. Our computation reproduces the same result as the one in \href{https://arxiv.org/abs/2203.04270}{arXiv:2203.04270} derived from a different approach.

\end{abstract}

\baselineskip 18pt
\thispagestyle{empty}

\newpage

\tableofcontents

\section{Introduction}

The study of diffeomorphism-invariant observables is a longstanding issue in the canonical formulation of gravitational theories \cite{Arnowitt:1962hi,Dirac:1958sc}. A prominent example of a diffeomorphism-invariant observable is the Hubeny-Rangamani-Takayanagi (HRT) area \cite{Hubeny:2007xt}, which is a covariant version of Ryu-Takayanagi (RT) formula \cite{Ryu:2006bv,Ryu:2006ef}.

As a diffeomorphism-invariant observable, the HRT area generates a Hamiltonian flow known as the HRT area flow \cite{Kaplan:2022orm}. It was suggested in \cite{Kaplan:2022orm} that the HRT area flow induces a boundary-condition-preserving (BCP) kink transformation when acting on initial data on a Cauchy slice that runs through the HRT surface. In the holographic context, the dual action of the HRT area flow is manifested through the boundary modular Hamiltonian $K_{bdy}$, since the HRT area can be regarded as an operator in quantum gravity obeying the Jafferis–Lewkowycz–Maldacena–Suh (JLMS) formula \cite{Jafferis:2014lza,Jafferis:2015del,Dong:2016eik}
\be K_{bdy}=\frac{A_{\text{HRT}}(\gamma_R)}{4G}+K_{bulk}, \ee
which creates a quantum-level connection between boundary CFT state and the bulk state enclosed by the HRT surface.

Further analysis in \cite{Kaplan:2022orm} examined how the HRT area flow acts on other observables, such as the boundary stress tensor and other HRT areas. However, the analysis of the boundary stress tensor in \cite{Kaplan:2022orm} proceeded in a dual manner: instead of analyzing the HRT area flow directly, they considered the conformal transformation generated by the boundary modular Hamiltonian—holographically dual to the HRT area. Our work seeks to fill this gap by the canonical formalism of gravity.

When applying the canonical formalism of gravity, a crucial step is to represent a given observable $O$ in terms of the set of initial data on a Cauchy slice, e.g., the induced metric $\sigma_{ab}$ and the extrinsic curvature $K_{ab}$ on the Cauchy surface in pure Einstein-Hilbert gravity. Then we can represent the action of the HRT area flow on the observable $O$ by chain rule
\be \D O=\int d\mathbf{x} \D\sigma_{ab}(t_0,\mathbf{x}) \frac{\delta O}{\delta \sigma_{ab}(t_0,\mathbf{x})}+\int d\mathbf{x} \D K_{ab}(t_0,\mathbf{x}) \frac{\delta O}{\delta K_{ab}(t_0,\mathbf{x})}, \ee
where $\D$ denotes the action of the HRT area flow, e.g., $\D\sigma_{ab}$ and $\D K_{ab}$ denote the action of the HRT area flow on the set of initial data which are the known results in \cite{Kaplan:2022orm}. 

Therefore, a key step toward our goal is to represent the boundary stress tensor $T^{(0)}_{ab}$ in terms of the initial data on the Cauchy surface. To this end, we need to introduce the relevant holographic renormalization procedure \cite{Henningson:1998gx,Balasubramanian:1999re,Bianchi:2001kw,deHaro:2000vlm}. Based on the asymptotic boundary conditions \footnote{The reason for this choice can refer to \cite{Wang:2023jfr}.} appearing in \cite{Wang:2023jfr}, we perform an off-shell holographic renormalization such that the renormalized action is finite for any given configuration satisfying the asymptotic boundary conditions. We refer the reader to our other paper \cite{ongoing1} for a more detailed treatment of the holographic renormalization. With the holographic renormalization at hand, we rederive the asymptotic form of the boundary stress tensor in terms of the induced metric and extrinsic curvature on the cut-off surface, in agreement with the result obtained in \cite{Henningson:1998gx,Balasubramanian:1999re,Bianchi:2001kw,deHaro:2000vlm}.

Solving the projection equations on the Cauchy surface and equations of motion then allow us to represent the boundary stress tensor in terms of the initial data on the Cauchy surface. Equivalently, this procedure amounts to solving for the lapse function $N$ and shift vector $\hat{\beta}$ appearing in the foliation of spacetime in terms of the initial data on the Cauchy surface. It is worth emphasizing that an appropriate choice of asymptotic boundary conditions can greatly simplify the analysis, for instance by reducing the couplings among certain variables.

This initial-data representation of the boundary stress tensor allows us not only to reproduce the results in \cite{Kaplan:2022orm}, but also to investigate several related issues. It is worth noting that Bousso et al.\cite{Bousso:2020yxi} got a stress tensor shock proceeding from a kink transformation, which is different from the BCP kink transformation generated by the HRT area flow. The former accompanies a kink transformation on the asymptotic boundary, while the latter preserves the boundary conditions. Here the stress tensor shock is viewed as the change between the boundary stress tensor with respect to kink transformed initial data in a tilded coordinates where the Cauchy surface can be viewed as a ``flat" surface\footnote{Here the word ``flat" denotes that the Cauchy surface is located at $\widetilde{t}=0$ in tilded coordinates.} and the one with respect to the initial data in initial coordinates. We also reproduce this stress tensor shock in a natural way.

In addition, we discuss other observables. An example is the twist along a geodesic \cite{Wang:2023jfr} in AdS$_3$, which is conjugate to the HRT area in some sense. We examine the action of the twist flow (the Hamiltonian flow generated by the twist) on the boundary stress tensor, which is consistent with the result in \cite{Wang:2023jfr}. The conjugate nature of these observables is also manifest in how the boundary stress tensor evolves under the corresponding Hamiltonian flow.

We expect that our approach may be applicable to a broader class of gravitational systems. For example, in more general gravitational theories containing higher-derivative interactions, the area functional is generalized to a geometric entropy functional $\sigma(\gamma_R)$ \cite{Dong:2013qoa,Camps:2013zua,Miao:2014nxa,Dong:2019piw}, obtained from the replica method in the bulk and capturing the full set of higher-curvature contributions. The surface $\gamma_R$ is determined by extremizing this generalized functional, which typically allows for a finite conical angle at $\gamma_R$ and reduces to the usual RT formula in the Einstein limit. There has been growing interest in studying the Hamiltonian flow of such geometric entropies \cite{Dong:2025orj} due to its importance in gravitational RG flow \cite{Dong:2025crd}.

The plan for the rest of the paper is as follows. In section \ref{sec:HRT}, we review the HRT area and its properties, especially the HRT area flow. In section \ref{sec:BST}, we define the boundary stress tensor in AdS$_3$ and represent the boundary stress tensor in terms of the initial data on the Cauchy surface. In section \ref{sec:HRTaction}, we study the action of the HRT area on the boundary stress tensor. In section \ref{sec:discussion}, we finish with conclusions and discussions.

\section{The HRT area}\label{sec:HRT}

In this section, we review the HRT area and its properties, especially the Hamiltonian flow it generates. The HRT area $A_{\text{HRT}}$ has a clear geometric significance. Namely, given a boundary region $R$, we can construct a series of codimension-$2$ surface in the bulk anchored to the boundary $\partial R$ of $R$ and find an extremal surface $\gamma_R$ with minimal area $A_{\text{HRT}}[\gamma_R]$. Here we denote the extremal surface $\gamma_R$ as the HRT surface and its area $A_{\text{HRT}}[\gamma_R]$ as the HRT area. Specifically in AdS$_3$, the HRT surface reduces to a geodesic and the HRT area reduces to the length of the geodesic. The HRT area establishes a link between the entanglement entropy of a region $R$ in the boundary CFT and geometric entropy of the HRT surface $\gamma_R$ anchored to $\partial R$ in the bulk AdS by
\be\label{EE} S(R)=\frac{A_{\text{HRT}}(\gamma_R)}{4G}. \ee

It has been revealed that the HRT area has various properties, mainly owing to its two features: as a kind of entropy with coefficient $1/(4G)$ and as an operator/observable in gravity. When viewed as the geometric entropy, it always appears in quantum entanglement in the dual CFT, even with higher derivative terms included \cite{Dong:2013qoa}, or appears in generalized entropy for general subregions in quantum gravity \cite{Jensen:2023yxy} and for the case of no dual CFT \cite{Chandrasekaran:2022eqq,Colafranceschi:2023moh}. When viewed as an operator/observable in gravity, it plays a vital role in the decomposition of the Hilbert space for gravity and constitutes an algebra in HRT surface networks\cite{Held:2024bgs}. Moreover, since it can be viewed as an operator/observable, we can act\footnote{Here the action of an operator/observable can be viewed that one acts it on the covariant phase space of classical solutions.} it on the gravitational system and analyze the associated evolution. It also takes a leading role in the canonical formalism of gravity. Therefore, in this work, we mainly study the action of the HRT area flow with canonical formalism of gravity.

We now give a precise representation of the action of the HRT area flow with canonical formalism of gravity. We first reformulate the gravitational system by a foliation of spacetime with a set of Cauchy surfaces containing the HRT surface. We regard a Cauchy surface as a dynamical object and take the induced metric $\sigma_{ab}$ and the extrinsic curvature $K_{ab}$ on the Cauchy surface as the set of initial data. Following the setup in \cite{Kaplan:2022orm}, we present the action of the HRT area flow on the initial data as
\begin{align}\label{HRTflowonid}
\Delta\sigma_{ab}=&\lambda\{\sigma_{ab},A_{\text{HRT}}\}+o(\lambda)\notag\\
\Delta K_{ab}=&\lambda\{K_{ab},A_{\text{HRT}}\}+o(\lambda).
\end{align}
where $\lambda$ is an infinitesimal parameter. For simplicity, in the following we will use the Poisson bracket for the bracket appearing in (\ref{HRTflowonid}).

It has been suggested in \cite{Kaplan:2022orm} that the action of the HRT area flow on the initial data (\ref{HRTflowonid}) yields a geometric interpretation, which is a boundary-condition-preserving kink transformation. We revisit it in one of our works\cite{Wang:2023jfr}, where we represent the action of the HRT area flow on the initial data (\ref{HRTflowonid}) as
\begin{align}\label{exHRTflowonid}
\Delta\sigma_{ab}=&o(\lambda)\notag\\
\Delta K_{ab}=&8\pi G\lambda\delta(\rho)n_an_b+o(\lambda),
\end{align}
where $\rho$ is a scalar field with $\rho=0$ on the HRT surface, $n_a$ is a covector normal to the HRT surface on the Cauchy surface. For convenience, we will stick with this setting for the rest of this paper. From (\ref{exHRTflowonid}), we can see that this action does not affect the initial data on both sides of the HRT surface, while it produces a change of extrinsic curvature when crossing the HRT surface. And there are no boundary conditions broken, especially the asymptotic boundary of AdS. So the action of the HRT area flow on the initial data can be interpreted as a BCP kink transformation from a pushforward perspective. It is a bit different from the ``kink transform" defined in \cite{Bousso:2020yxi}. We will discuss this difference in detail in section \ref{sec:discussion} and apply our approach to this issue.

Given the action of the HRT area flow on the initial data (\ref{exHRTflowonid}), we can, in principle, compute the action of the HRT area flow on any diffeomorphism-invariant observable $O$ in gravity. We just need to represent the observable $O$ in terms of the initial data on the Cauchy surface
\be\label{Oid} O\equiv O(\sigma_{ab},K_{ab}), \ee
and use the chain rule to represent the action of the HRT area flow on $O$ as
\be\label{chainrule} \D O=\int d\mathbf{x} \D\sigma_{ab}(t_0,\mathbf{x}) \frac{\delta O}{\delta \sigma_{ab}(t_0,\mathbf{x})}+\int d\mathbf{x} \D K_{ab}(t_0,\mathbf{x}) \frac{\delta O}{\delta K_{ab}(t_0,\mathbf{x})}. \ee
Generally speaking, it is a challenge to reach the first step (\ref{Oid}). Once the first step (\ref{Oid}) is completed, the second step (\ref{chainrule}) is straightforward.

In this work, we are mainly interested in the boundary stress tensor $T^{(0)}_{ab}$ which is a crucial quantity for vacuum Poincare AdS. And we expect to revisit the action of the HRT area flow on the boundary stress tensor with canonical formalism of gravity, which is a loss in \cite{Kaplan:2022orm}. It is a further support for the statement that the action of the HRT area flow generates a kink transformation in pure AdS$_3$ gravity. In section \ref{sec:BST}, we will represent the boundary stress tensor in terms of the initial data, under proper asymptotic boundary conditions. Then we will take a straightforward computation of the action of the HRT area flow on the boundary stress tensor and reproduce a specific interpretation--a stress tensor shock in section \ref{sec:HRTaction}.

\section{The boundary stress tensor}\label{sec:BST}

In this section, we introduce the boundary stress tensor and represent it in terms of the initial data on the Cauchy surfaces in AdS$_3$.

\subsection{The setup}

We begin by presenting the setup for AdS$_3$, together with the relevant holographic renormalization and the construction of the boundary stress tensor.

Note that the conformal compactification of AdS$_3$ yields a compact manifold with a timelike asymptotic (conformal) boundary. We denote the bulk region by $\mathcal{M}$, the asymptotic boundary by $\Gamma$, and the action of pure AdS$_3$ gravity by
\be\label{action0} S=\frac{1}{16\pi G}\int_{\mathcal{M}}\sqrt{-g}(R+2)+\frac{1}{8\pi G}\int_\Gamma\sqrt{-\gamma}\mathcal{K}-\frac{1}{8\pi G}\int_\Gamma\sqrt{-\gamma}, \ee
where the first term is the Einstein-Hilbert action $S_{\text{EH}}$, the second term is the Gibbons–Hawking–York boundary term $S_{\text{GHY}}$ for a well-defined variational principle, the third term is a counter term $S_{ct}$ for a finite boundary stress tensor. Here we denote the induced metric of asymptotic boundary by $\gamma_{ab}$ and the extrinsic curvature of the asymptotic boundary by $\mathcal{K}$.

We need to introduce the holographic renormalization to obtain a regularized version of spacetime and the action (\ref{action0}). More precisely, we introduce an IR cutoff surface labeled by a parameter $\epsilon$\footnote{In Poincare coordinates, we usually take the radial-like coordinate $z=\epsilon$. In this paper we study the AdS$_3$ in Poincare patch. If you would like to transforms to the global AdS$_3$, the identification between the Poincare AdS$_3$ boundary and the global AdS$_3$ boundary should be noticed.\cite{Bayona:2005nq}}. We adopt the following asymptotic boundary conditions\footnote{Although there is the most general AdS$_3$ asymptotic boundary conditions in \cite{Grumiller:2016pqb}, one could only adopt the most appropriate asymptotic boundary condition for the theory under consideration. It is equivalent to choosing a proper representation of asymptotic symmetries in some sense.}
\begin{align}\label{ABC0}
	&g_{zz}=\frac{1}{z^2}+\mathcal{O}(z^0)\notag\\
	&g_{za}=\mathcal{O}(\frac{1}{z})\notag\\
	&g_{ab}=\frac{1}{z^2}g^{(0)}_{ab}+\mathcal{O}(z^0),
\end{align}
for the metric $g_{\mu\nu}$, where we have taken a general asymptotic boundary instead of a flat boundary such that we have a general asymptotic boundary metric $g^{(0)}_{ab}$ only depending on the boundary coordinate $x^a$. Here we choose a weaker asymptotic boundary condition in \cite{Wang:2023jfr} rather than the widely used one in \cite{Brown:1986nw}. We collect asymptotic behaviors of related quantities in Appendix \ref{appendix:asy}.

We then denote the regularized version of the asymptotic boundary by $\Gamma_\epsilon$, the regularized version of action (\ref{action0}) by
\be\label{action1} S_\epsilon=\frac{1}{16\pi G}\int_{\mathcal{M}_\epsilon}\sqrt{-g}(R+2)+\frac{1}{8\pi G}\int_{\Gamma_\epsilon}\sqrt{-\gamma}(\mathcal{K}-1)+F(\epsilon), \ee
where $F(\epsilon)$ collects the counter terms supported on $\Gamma_\epsilon$. A more detailed analysis in \cite{ongoing1} shows that
\be F(\epsilon)=o(\epsilon^s\log\epsilon)+\mathcal{O}(\epsilon^2\log\epsilon), \ee 
for some positive real number $s$. Therefore we can denote the action (\ref{action0}) as a limit of the regularized version (\ref{action1})
\be S=\lim\limits_{\epsilon\to 0}S_\epsilon. \ee
 
We then define a given physical quantity $O^{(0)}$ supported at the asymptotic boundary $\Gamma$ by the limit of a regularized version of $O_\epsilon$ supported at $\Gamma_\epsilon$
\be O^{(0)}\Big|_\Gamma=\lim\limits_{\epsilon\to 0}O_\epsilon\Big|_{\Gamma_\epsilon}. \ee

We now introduce the boundary stress tensor. The boundary stress tensor is a quasilocal stress tensor (Brown-York tensor\cite{Brown:1992br}) defined on the boundary of a given spacetime region, which can be regarded as a remedy for the fact that we are unable to define a local energy-momentum tensor for the gravitational field in a natural way. Here we define it by
\be\label{def bst}
T^{\mu\nu}=-\frac{4\pi}{\sqrt{-\gamma}}\frac{\delta S}{\delta\gamma_{\mu\nu}},
\ee
where the prefactor $-\frac{4\pi}{\sqrt{-\gamma}}$ is determined by the charge associated with the asymptotic symmetries in \cite{Wang:2023jfr} and consistent with the boundary stress tensor under the Fefferman-Graham gauge. 

The definition of the boundary stress tensor (\ref{def bst}) implies a boundary stress tensor represented in terms of the induced metric $\gamma_{ab}$ and the extrinsic curvature $\mathcal{K}_{ab}$ on the asymptotic boundary. To achieve our goal, we will derive an asymptotic form of the boundary stress tensor from (\ref{ABC0}). Based on this asymptotic expression, we then solve the projection equation together with the bulk equations of motion, obtaining an initial data form of the boundary stress tensor\footnote{Geometrically, it lives on the intersection of the Cauchy surface and the asymptotic boundary.}.

\subsection{The asymptotic form of boundary stress tensor}

In this subsection, we derive an asymptotic form of boundary stress tensor based on the asymptotic boundary conditions (\ref{ABC0}).

Applying (\ref{action1}) on (\ref{def bst}) yields a widely used form of the boundary stress tensor \cite{Balasubramanian:1999re}
\be\label{bst1} T^{(0)}_{ab}=\lim\limits_{\epsilon\to 0}T_{ab}\Big|_{z=\epsilon}=\lim_{\epsilon\to 0}\frac{1}{4G}(\mathcal{K}_{ab}-\mathcal{K}\gamma_{ab}+\gamma_{ab})\Big|_{z=\epsilon}, \ee 
explicitly expressed with $\gamma_{ab}$ and $\mathcal{K}_{ab}$ on $\Gamma_\epsilon$.

Applying (\ref{ABCgamma}), (\ref{ABCKGamma}) and (\ref{ABCKG}) on (\ref{bst1}) yields an asymptotic form of the boundary stress tensor
\begin{align}\label{ABCbst}
	T^{(0)}_{ab}=\lim_{\epsilon\to 0}\frac{1}{4G}\Big[&\frac{1}{2}(g_{zz}-\frac{1}{z^2})g^{(0)}_{ab}-(g_{ab}-\frac{1}{z^2}g^{(0)}_{ab})-\frac{1}{2}z\partial_z(g_{ab}-\frac{1}{z^2}g^{(0)}_{ab})\notag\\
	&+g^{(0)cd}(g_{cd}-\frac{1}{z^2}g^{(0)}_{cd})g^{(0)}_{ab}+\frac{1}{2}zg^{(0)cd}\partial_z(g_{cd}-\frac{1}{z^2}g^{(0)}_{cd})g^{(0)}_{ab}\notag\\
	&+\frac{1}{2}z(D^{(0)}_ag_{zb}+D^{(0)}_bg_{za})-zg^{(0)cd}D^{(0)}_cg_{zd}g^{(0)}_{ab}\notag\\
	&-\frac{1}{2}z^2g^{(0)cd}g_{zc}g_{zd}g^{(0)}_{ab}+\mathcal{O}(z^2)\Big]\Big|_{z=\epsilon},
\end{align}
where we denote the covariant derivative associated with $g^{(0)}_{ab}$ by $D^{(0)}$.

\subsection{The initial-data form of boundary stress tensor}

In this subsection, we derive an initial-data form of boundary stress tensor. Namely, we represent the boundary stress tensor in terms of initial data on the Cauchy surface $\Sigma$.

Given the asymptotic form of the boundary stress tensor (\ref{ABCbst}), we introduce a foliation of spacetime with a set of Cauchy surfaces to develop a tie between this asymptotic form and initial data on the Cauchy surface. We collect the details of the foliation in Appendix \ref{appendix:Foliation}.

\begin{figure}[h]
	\begin{minipage}[b]{0.47\linewidth}
		\centering
		\subfloat[][A foliation of spacetime with a set of Cauchy surfaces labelled by a function $T$, with a regularized version of the bulk spacetime.]{\label{F1a}\includegraphics[width=0.95\linewidth]{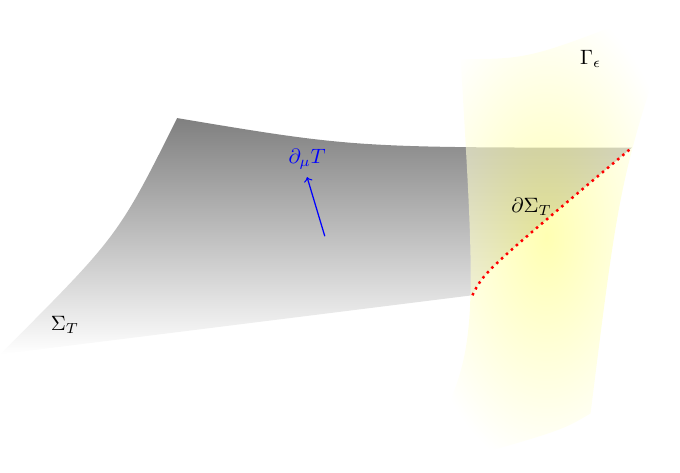}}
	\end{minipage}
	\begin{minipage}[b]{0.47\linewidth}
		\centering
		\subfloat[][A foliation of asymptotic boundary with a set of Cauchy surfaces labelled by a function $T^{(0)}$]{\label{F1b}\includegraphics[width=0.8\linewidth]{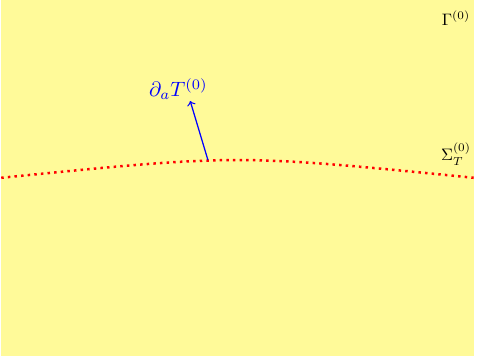}}
	\end{minipage}
	\caption{A match of foliation: (a) illustrates the foliation of bulk spacetime, where the gray surface denotes the Cauchy surface $\Sigma_T$ with a constant $T$, the yellow region denotes the IR cutoff surface $\Gamma_\epsilon$ and the red dotted line denotes the intersection between both surfaces;(b) illustrates the foliation of the asymptotic boundary $\Gamma^{(0)}$, where the red dotted line denotes the Cauchy surface $\Sigma^{(0)}$ with a constant $T^{(0)}$. We can match them by setting $T=T^{(0)}$.}\label{Fig:foliation}
\end{figure}

This foliation of bulk spacetime can manifestly induce a foliation of the asymptotic boundary if we determine how the Cauchy surface extends to asymptotic boundary, more precisely, how the function $T$ labelling the Cauchy surface $\Sigma$ and its regularized boundary $\partial\Sigma$ matches with the function $T^{(0)}$ labelling the Cauchy surface $\Sigma^{(0)}$ in the asymptotic boundary, as shown in Fig.\ref{Fig:foliation}. 

In this sense, we can represent the normal and tangential projections of the boundary stress tensor in terms of the normal and tangential projections of the metric $g_{\mu\nu}$ (or an equivalent and more familiar form, the lapse function $N$ and the shift vector $\hat{\beta}$). More precisely, based on the twin foliation in Appendix \ref{appendix:Foliation}, we first project the boundary stress tensor (\ref{ABCbst}) into the energy surface density $u^{(0)a}u^{(0)b}T^{(0)}_{ab}$, the momentum surface density $u^{(0)a}h^{(0)b}_cT^{(0)}_{ab}$ and the spatial stress $h^{(0)a}_ch^{(0)b}_dT^{(0)}_{ab}$
\begin{align}\label{Pbst}
	u^{(0)m}u^{(0)n}T^{(0)}_{mn}=\lim_{\epsilon\to 0}\frac{1}{4G}\Big[&-\frac{1}{2}(g_{zz}-\frac{1}{z^2})-h^{(0)ab}(g_{ab}-\frac{1}{z^2}g^{(0)}_{ab})-\frac{1}{2}zh^{(0)ab}\partial_z(g_{ab}-\frac{1}{z^2}g^{(0)}_{ab})\notag\\
	&+zh^{(0)ab}D^{(0)}_a(h^{(0)c}_bg_{zc})-zK^{(0)}u^{(0)a}g_{za}\notag\\
	&+\frac{1}{2}z^2h^{(0)ab}g_{za}g_{zb}-\frac{1}{2}z^2(u^{(0)a}g_{za})^2+\mathcal{O}(z^2)\Big]\Big|_{z=\epsilon}\notag\\
	h^{(0)m}_au^{(0)n}T^{(0)}_{mn}=\lim_{\epsilon\to 0}\frac{1}{4G}\Big[&-h^{(0)b}_au^{(0)c}(g_{bc}-\frac{1}{z^2}g^{(0)}_{bc})-\frac{1}{2}zh^{(0)b}_au^{(0)c}\partial_z(g_{bc}-\frac{1}{z^2}g^{(0)}_{bc})\notag\\
	&-\frac{1}{2}zK^{(0)}h^{(0)b}_ag_{zb}+\frac{1}{2}zh^{(0)b}_au^{(0)c}D^{(0)}_c(h^{(0)d}_bg_{zd})\notag\\
	&+\frac{1}{2}zh^{(0)b}_aD^{(0)}_b(u^{(0)c}g_{zc})-\frac{1}{2}za^{(0)}_au^{(0)b}g_{zb}+\mathcal{O}(z^2)\Big]\Big|_{z=\epsilon}\notag\\
	h^{(0)m}_ah^{(0)n}_bT^{(0)}_{mn}=\lim_{\epsilon\to 0}\frac{1}{4G}\Big[&\Big(\frac{1}{2}(g_{zz}-\frac{1}{z^2})-u^{(0)c}u^{(0)d}(g_{cd}-\frac{1}{z^2}g^{(0)}_{cd})-\frac{1}{2}zu^{(0)c}u^{(0)d}\partial_z(g_{cd}-\frac{1}{z^2}g^{(0)}_{cd})\notag\\
	&-zh^{(0)cd}a^{(0)}_cg_{zd}+zu^{(0)c}D^{(0)}_c(u^{(0)d}g_{zd})\notag\\
	&-\frac{1}{2}z^2h^{(0)cd}g_{zc}g_{zd}+\frac{1}{2}z^2(u^{(0)c}g_{zc})^2\Big)h^{(0)}_{ab}+\mathcal{O}(z^2)\Big]\Big|_{z=\epsilon},
\end{align}
where we have used (\ref{unitu}), (\ref{hab}), (\ref{a0}), (\ref{K0ab}), (\ref{K0}) and (\ref{dim1}). 

We can see that the projection of the metric, e.g., $h^{(0)b}_ag_{zb}$, appears in (\ref{Pbst}). To relate them with initial data on the Cauchy surface $\Sigma$, we consider the projection of the set of initial data
\begin{align}\label{Psigma}
	\sigma_{zz}=&\frac{1}{z^2}+(g_{zz}-\frac{1}{z^2})\notag\\
	h^{(0)m}_a\sigma_{zm}=&h^{(0)b}_ag_{zb}\notag\\
	h^{(0)m}_ah^{(0)n}_b\sigma_{mn}=&\frac{1}{z^2}h^{(0)}_{ab}+h^{(0)cd}(g_{cd}-\frac{1}{z^2}g^{(0)}_{cd})h^{(0)}_{ab}+\mathcal{O}(z^2),
\end{align}
and
\begin{align}\label{PKab}
	K_{zz}=&-u^{(0)a}g_{za}-zu^{(0)a}\partial_zg_{za}+\mathcal{O}(z)\notag\\
	h^{(0)m}_aK_{zm}=&-h^{(0)b}_au^{(0)c}(g_{bc}-\frac{1}{z^2}g^{(0)}_{bc})-\frac{1}{2}zh^{(0)b}_au^{(0)c}\partial_z(g_{bc}-\frac{1}{z^2}g^{(0)}_{bc})+\frac{1}{2}zK^{(0)}h^{(0)b}_ag_{zb}\notag\\
	&-\frac{1}{2}zh^{(0)b}_aD^{(0)}_b(u^{(0)c}g_{zc})+\frac{1}{2}zh^{(0)b}_au^{(0)c}D^{(0)}_c(h^{(0)d}_bg_{zd})-\frac{1}{2}za^{(0)}_au^{(0)b}g_{zb}\notag\\
	&+z^2h^{(0)b}_au^{(0)c}g_{zb}g_{zc}+\mathcal{O}(z^2)\notag\\
	h^{(0)m}_ah^{(0)n}_bK_{mn}=&\frac{1}{z}K^{(0)}_{ab}+u^{(0)c}g_{zc}h^{(0)}_{ab}+\mathcal{O}(z),
\end{align}
where we have used the asymptotic behaviors (\ref{ABCsigma_mu^nu}) and (\ref{ABCK_munu}). Solving (\ref{Psigma}) and (\ref{PKab}), and applying the solutions on (\ref{Pbst}) yields the initial-data form of the boundary stress tensor
\begin{align}\label{Pbstid}
	u^{(0)m}u^{(0)n}T^{(0)}_{mn}=\lim_{\epsilon\to 0}\frac{1}{4G}\Big[&-\frac{1}{2}(\sigma_{zz}-\frac{1}{z^2})-(h^{(0)ab}\sigma_{ab}-\frac{1}{z^2})-\frac{1}{2}z\partial_z(h^{(0)ab}\sigma_{ab}-\frac{1}{z^2})\notag\\
	&+zh^{(0)ab}D^{(0)}_a(h^{(0)c}_b\sigma_{zc})+\frac{1}{2}(K^{(0)})^2\notag\\
	&-\frac{1}{2}z^2(h^{(0)ab}K_{ab})^2+\frac{1}{2}z^2h^{(0)ab}\sigma_{za}\sigma_{zb}+\mathcal{O}(z^2)\Big]\Big|_{z=\epsilon}\notag\\
	h^{(0)m}_au^{(0)n}T^{(0)}_{mn}=\lim_{\epsilon\to 0}\frac{1}{4G}\Big[&h^{(0)b}_aK_{zb}+h^{(0)b}_aD^{(0)}_b(zh^{(0)cd}K_{cd}-K^{(0)})-z^2h^{(0)b}_a\sigma_{zb}h^{(0)cd}K_{cd}+\mathcal{O}(z^2)\Big]\Big|_{z=\epsilon}\notag\\
	h^{(0)m}_ah^{(0)n}_bT^{(0)}_{mn}=\lim_{\epsilon\to 0}\frac{1}{4G}\Big[&\Big(-\frac{1}{2}(\sigma_{zz}-\frac{1}{z^2})-(h^{(0)cd}\sigma_{cd}-\frac{1}{z^2})-\frac{1}{2}z\partial_z(h^{(0)cd}\sigma_{cd}-\frac{1}{z^2})\notag\\
	&+zh^{(0)cd}D^{(0)}_c(h^{(0)f}_d\sigma_{zf})-\frac{1}{2}R^{(0)}+\frac{1}{2}(K^{(0)})^2\notag\\
	&-\frac{1}{2}z^2(h^{(0)cd}K_{cd})^2+\frac{1}{2}z^2h^{(0)cd}\sigma_{zc}\sigma_{zd}\Big)h^{(0)}_{ab}+\mathcal{O}(z^2)\Big]\Big|_{z=\epsilon},
\end{align}
where in deriving $h^{(0)m}_ah^{(0)n}_bT^{(0)}_{mn}$, for simplicity, we have used the trace condition\footnote{Here we also have implicitly used the equations of motion since we use which in deriving the trace condition for $T^{(0)}_{ab}$.} for $T^{(0)}_{ab}$
\be
g^{(0)ab}T^{(0)}_{ab}=(h^{(0)ab}-u^{(0)a}u^{(0)b})T^{(0)}_{ab}=\frac{1}{4G}\Big(-\frac{1}{2}R^{(0)}\Big).
\ee
Here we should remind readers that when we derive (\ref{Pbstid}), we need to solve the projection of $g_{\mu\nu}$ with respect to the projection of the initial data by solving (\ref{Psigma}) and (\ref{PKab}), e.g., 
\begin{align}
	(g_{zz}-\frac{1}{z^2})=&(\sigma_{zz}-\frac{1}{z^2})\notag\\
	h^{(0)b}_ag_{zb}=&h^{(0)m}_a\sigma_{zm}\notag\\
	h^{(0)ab}(g_{ab}-\frac{1}{z^2}g^{(0)}_{ab})=&(h^{(0)ab}\sigma_{ab}-\frac{1}{z^2})\notag\\
	u^{(0)a}g_{za}=&h^{(0)ab}K_{ab}-\frac{1}{z}K^{(0)}+\mathcal{O}(z),
\end{align}
where these solutions actually are dependent of the asymptotic conditions for the metric $g_{\mu\nu}$ (\ref{ABC0}). One can also translate this procedure into the language with the lapse function $N$ and shift vector $\hat{\beta}$.

So far, we have represented the boundary stress tensor in terms of the initial data in (\ref{Pbstid}). Then we can further study the action of the HRT area flow on the boundary stress tensor with canonical formalism of gravity.


\section{The action of HRT area flow on the boundary stress tensor}\label{sec:HRTaction}
In this section, we revisit the action of the HRT area flow on the boundary stress tensor with canonical formalism of gravity and give an interpretation--the stress tensor shock. 

Given the previous setup, the computation reduces to inserting the initial-form of boundary stress tensor (\ref{Pbstid}) into the position of the observable $O$ in (\ref{chainrule}), and subsequently applying (\ref{exHRTflowonid}) to arrive at our goal.

For comparison with the result in \cite{Kaplan:2022orm}, we specialize to the case of a flat asymptotic boundary such that $g^{(0)}_{ab}$ reduces to the Minkowski metric
\be\label{flatbdy} g^{(0)}_{ab}=\eta_{ab}. \ee
And we choose $(t,x)$ coordinates associated with the flat metric (\ref{flatbdy}) such that
\be\label{u0flat} u^{(0)}_t=1,\quad u^{(0)}_x=0. \ee

Applying (\ref{flatbdy}), (\ref{u0flat}) and (\ref{hab}) on (\ref{Pbstid}) yields the components of the initial-data form of the boundary stress tensor
\begin{align}\label{bstflat}
T^{(0)}_{tt}=\lim\limits_{\epsilon\to 0}\frac{1}{4G}\Big[&-\frac{1}{2}(\sigma_{zz}-\frac{1}{z^2})-(\sigma_{xx}-\frac{1}{z^2})-\frac{1}{2}z\partial_z(\sigma_{xx}-\frac{1}{z^2})\notag\\
&+\frac{1}{2}z^2\sigma^2_{zx}-\frac{1}{2}z^2K^2_{xx}+z\partial_x\sigma_{zx}+\mathcal{O}(z^2)\Big]\Big|_{z=\epsilon}\notag\\
T^{(0)}_{tx}=\lim\limits_{\epsilon\to 0}\frac{1}{4G}\Big[&K_{zx}+z\partial_xK_{xx}-z^2K_{xx}\sigma_{zx}+\mathcal{O}(z^2)\Big]\Big|_{z=\epsilon}\notag\\
T^{(0)}_{xx}=\lim\limits_{\epsilon\to 0}\frac{1}{4G}\Big[&-\frac{1}{2}(\sigma_{zz}-\frac{1}{z^2})-(\sigma_{xx}-\frac{1}{z^2})-\frac{1}{2}z\partial_z(\sigma_{xx}-\frac{1}{z^2})\notag\\
&+\frac{1}{2}z^2\sigma^2_{zx}-\frac{1}{2}z^2K^2_{xx}+z\partial_x\sigma_{zx}+\mathcal{O}(z^2)\Big]\Big|_{z=\epsilon},
\end{align}
which are expressed explicitly in terms of the exponents of the initial data on the Cauchy surface $\Sigma_T$. 

Applying (\ref{chainrule}) on (\ref{bstflat}) yields the initial-form of the action of the HRT area flow on the boundary stress tensor
\begin{align}\label{Dbstid}
\Delta T^{(0)}_{tt}=&\lim\limits_{\epsilon\to 0}\frac{1}{4G}\Big[-\frac{1}{2}\Delta\sigma_{zz}-\Delta\sigma_{xx}+z^2\sigma_{zx}\Delta\sigma_{zx}-z^2K_{xx}\Delta K_{xx}+z\partial_x\Delta\sigma_{zx}\Big]\Big|_{z=\epsilon}\notag\\
\Delta T^{(0)}_{tx}=&\lim\limits_{\epsilon\to 0}\frac{1}{4G}\Big[\Delta K_{zx}+z\partial_x\Delta K_{xx}-z^2K_{xx}\Delta\sigma_{zx}-z^2\sigma_{zx}\Delta K_{xx}\Big]\Big|_{z=\epsilon}\notag\\
\Delta T^{(0)}_{xx}=&\lim\limits_{\epsilon\to 0}\frac{1}{4G}\Big[-\frac{1}{2}\Delta\sigma_{zz}-\Delta\sigma_{xx}+z^2\sigma_{zx}\Delta\sigma_{zx}-z^2K_{xx}\Delta K_{xx}+z\partial_x\Delta\sigma_{zx}\Big]\Big|_{z=\epsilon},
\end{align}      
where we denote the action of the HRT area flow by the symbol $\Delta$. Here $\Delta\sigma_{ab}$ and $\Delta K_{ab}$ are given in (\ref{exHRTflowonid}) following the conventions in \cite{Wang:2023jfr}.

\begin{figure}[h]
      \begin{minipage}[b]{0.6\linewidth}
      \centering
      \subfloat[][The action of the HRT area on the initial data in the case of planar asymptotic boundary, with a regularized version of the bulk spacetime. It can be represented as a kink transformation.]{\label{F2a}\includegraphics[width=0.95\linewidth]{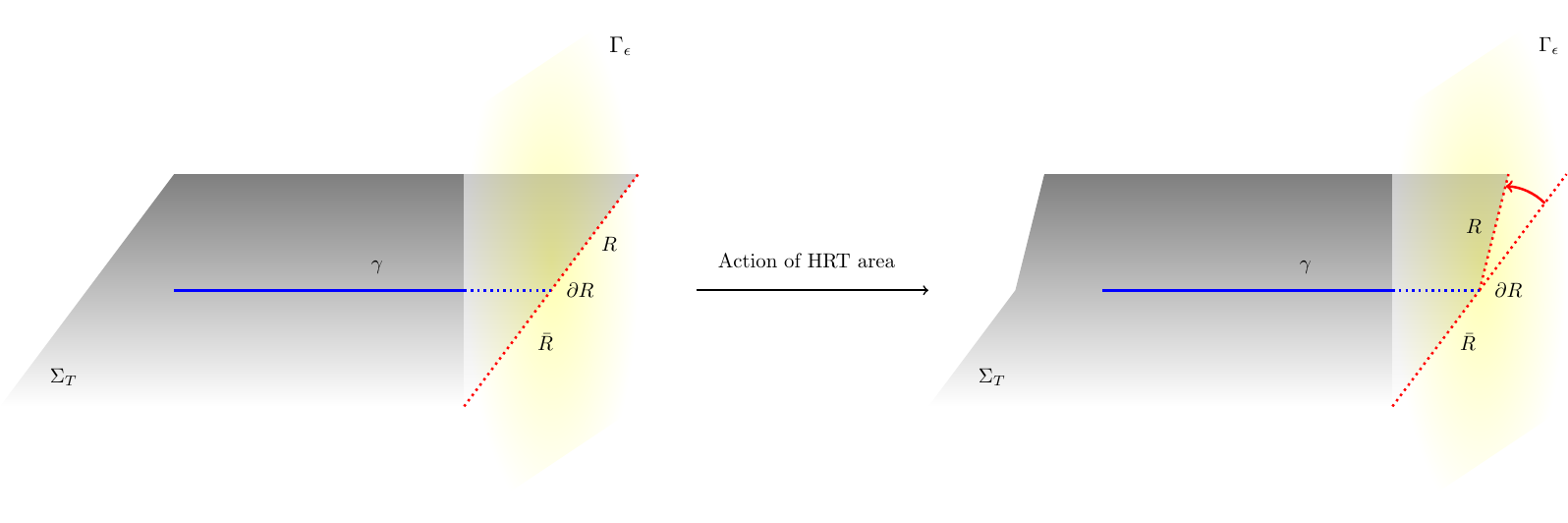}}
      \end{minipage}
      \begin{minipage}[b]{0.35\linewidth}
      \centering
      \subfloat[][The action of the HRT area preserves the asymptotic boundary.]{\label{F2b}\includegraphics[width=0.8\linewidth]{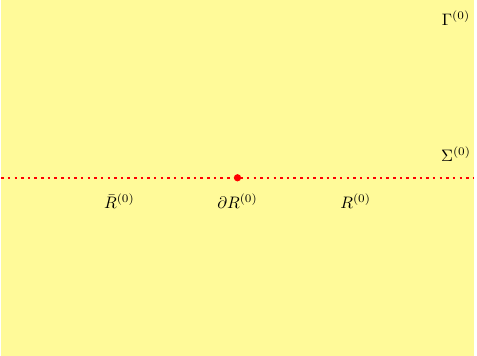}}
      \end{minipage}
     \caption{An interpretation of the action of the HRT area flow on the initial data in the case of planar asymptotic boundary: (a) illustrates the kink transformation of Cauchy surface in bulk spacetime, where the HRT area flow induces a relative boost between the left and right sides of the Cauchy surface $\Sigma_T$;(b) illustrates the preservation of the Cauchy surface $\Sigma^{(0)}$ on the asymptotic boundary $\Gamma^{(0)}$ under the action of the HRT area flow (Or precisely the action of dual Modular Hamiltonian). Before the action of the HRT area flow, we have matched them by setting $T=T^{(0)}$.}\label{Fig:actionflat}
\end{figure}

To give (\ref{Dbstid}) a clearer physical meaning and compare with the results in \cite{Kaplan:2022orm}, we further consider a simple interpretation in \cite{Kaplan:2022orm}, shown in Fig.\ref{Fig:actionflat}. More precisely, we choose the boundary region $R$ to be a half line with $x\in[0,\infty)$ at $t=0$ such that the HRT surface is the bulk geodesic parameterized by
\begin{align}\label{line}
z=&f(s)\notag\\
t=&t_0\notag\\
x=&x_0,
\end{align}
in the region near the asymptotic boundary, where $f(s)$ is an arbitrary smooth function with respect to an affine parameter $s$. 

In this picture, we can set the scalar field $\rho$ in (\ref{exHRTflowonid}) as
\be\label{rho} \rho=(\frac{1}{z}+\mathcal{O}(z))(x_0-x), \ee
and $n_a$ in (\ref{exHRTflowonid}) as
\begin{align}\label{n_a}
n_z=&(-\frac{1}{z^2}+\mathcal{O}(z^0))(x_0-x)\notag\\
n_x=&-\frac{1}{z}+\mathcal{O}(z),
\end{align}
where $\rho$ and $n_a$ satisfy the following relation
\be n_a|_\gamma=\partial_a\rho|_\gamma \ee
on the geodesic (the HRT surface) $\gamma$. In addition, we also introduce a vector field $\hat{e}$ by
\begin{align}\label{e_a}
	e_z=&\frac{1}{z}+\mathcal{O}(z)\notag\\
	e_x=&(-\frac{1}{z^2}+\mathcal{O}(z^0))(x_0-x),
\end{align}
which holds the orthogonality between $\hat{n}$ and $\hat{e}$.
Here $\hat{n}$ and $\hat{e}$ can be viewed as an extension of the normal vector and the tangent vector to the geodesic to the whole Cauchy surface. From (\ref{rho}), we can further get $\delta(\rho)$ in (\ref{exHRTflowonid}) by
\begin{align}\label{delrho}
&\theta(\rho)=\theta(x_0-x)\notag\\
&\delta(\rho)=(z+\mathcal{O}(z^3))\delta(x_0-x),
\end{align}
where $\theta(\rho)$ is the Heaviside step function. So far we have obtained explicit expressions for all quantities appearing in (\ref{exHRTflowonid}).

Applying (\ref{n_a}) and (\ref{delrho}) on (\ref{exHRTflowonid}), we represent the action of the HRT area on the initial data (\ref{exHRTflowonid}) as
\begin{align}\label{actid}
\Delta\sigma_{ab}=&o(\lambda)\notag\\
\Delta K_{zz}=&o(\lambda)\notag\\
\Delta K_{zx}=&o(\lambda)\notag\\
\Delta K_{xx}=&8\pi G\lambda\Big(\frac{1}{z}+\mathcal{O}(z)\Big)\delta(x_0-x)+o(\lambda).
\end{align}

Applying (\ref{actid}) on (\ref{Dbstid}) yields an interpretation of initial-data form of the action of the HRT area flow on the boundary stress tensor
\begin{align}\label{actbst}
\Delta T^{(0)}_{tt}=&\lim\limits_{\epsilon\to 0}(-2\pi)\Big(\lambda zK_{xx}\delta(x_0-x)+o(\lambda)\Big)\Big|_{z=\epsilon}\notag\\
\Delta T^{(0)}_{tx}=&\lim\limits_{\epsilon\to 0}(-2\pi)\Big[\lambda\Big(\delta^\prime(x_0-x)+z\sigma_{zx}\delta(x_0-x)\Big)+o(\lambda)\Big]\Big|_{z=\epsilon}\notag\\
\Delta T^{(0)}_{xx}=&\lim\limits_{\epsilon\to 0}(-2\pi)\Big(\lambda zK_{xx}\delta(x_0-x)+o(\lambda)\Big)\Big|_{z=\epsilon}.
\end{align}
which, under the setup that the boundary region is a half line with $x\in[0,\infty)$ at $t=0$, yields the only non-trivial component\footnote{We clarify this in Appendix \ref{appendix:clarify}.}
\be\label{actbst3} \Delta T^{(0)}_{tx}=\lim\limits_{\epsilon\to 0}(-2\pi)\Big(\lambda\delta^\prime(x_0-x)+o(\lambda)\Big)\Big|_{z=\epsilon}, \ee
which is consistent with the results in \cite{Kaplan:2022orm} at linear order in $\lambda$, and we need a potential regulator if we consider a finite transformation. It also shows a stress tensor shock, which is similar to an energy "pulse" concentrated on the HRT surface.

We end this section with some comments on a little weird term $\delta^\prime(x_0-x)$ appearing in (\ref{actbst}) and (\ref{actbst3}). It makes sense if we just consider an infinitesimal transformation. Nevertheless, we need to consider a regularization for the kinked Cauchy surfaces $\Sigma_T$, shown in Fig.\ref{Fig:actionflat}, if we seek to perform a finite transformation. A solution to hiding this weird term is to take a new foliation of asymptotic boundary with a set of Cauchy surface $\widetilde{\Sigma}^{(0)}_{T}$ labelled by $\widetilde{T}^{(0)}$ and take a new matching $T=\widetilde{T}^{(0)}$, such that when taking a pullback to the initial foliation, we have
\begin{align}
\widetilde{T}^{(0)}=&T^{(0)}, \text{for region}\ R^{(0)}\notag\\
\widetilde{T}^{(0)}=&(1+\#\lambda)T^{(0)}, \text{for region}\ \bar{R}^{(0)}.
\end{align}
We can also take an additional coordinate transformation such that it the Cauchy surface $\Sigma_T$ and $\Sigma^{(0)}_{T}$ seem smooth rather than kinked in this new coordinate. And we can further apply our approach on this case to investigate the change of the boundary stress tensor. A similar situation has already been considered in \cite{Bousso:2020yxi}. We will revisit it with our approach in Discussion.

\section{Conclusion and discussion}\label{sec:discussion}

In this paper, we revisit the HRT area and the action of the HRT area flow on the boundary stress tensor with canonical formalism of gravity. We obtain an initial-data form of the boundary stress tensor (\ref{Pbstid}) under asymptotic conditions (\ref{ABC0}). We also reproduce the action of the HRT area flow on the boundary stress tensor for a half line boundary region, namely the stress tensor shock (\ref{actbst3}) in \cite{Kaplan:2022orm}. This provides a further support for the statement that the action of the HRT-area generates a kink transformation in the pure AdS$_3$ gravity.

We also expect to seek out more potential applications for our approach. Throughout the previous steps, we have favored abstract index notation over coordinate expressions whenever possible, thereby enabling a seamless generalization to higher dimensions. Moreover, we can examine the influence of the Hamiltonian flow generated by other diffeomorphism-invariant observables on the boundary stress tensor. To provide further support for the statement that the action of the HRT-area generates a kink transformation in pure AdS$_3$ gravity, we can consider another kink transformation, which is dual to the Connes Cocycle flow \cite{Bousso:2020yxi}.

\subsection{The action of the twist flow on the boundary stress tensor}

In this subsection, we discuss the action of the twist flow \cite{Wang:2023jfr} on the boundary stress tensor.

The twist along the geodesic \cite{Wang:2023jfr} might be conjugate to the HRT area in the decomposition of the Hilbert space in AdS$_3$ gravity. This conjugation might manifest in the action of their Hamiltonian flow. 

Analogous to our treatment of the HRT area, we start by listing the action of the twist flow on the initial data on the Cauchy surface
\begin{align}\label{twflowid}
&\Delta \sigma_{ab}(t_0,x)=-8\pi G \lambda (e_an_b+n_ae_b) \delta(\rho) +o(\lambda) \notag \\
&\Delta K_{ab}(t_0,x)=-8\pi G \lambda e^m n^n K_{mn} n_an_b \delta(\rho) +o(\lambda),
\end{align}
where $\D$ denotes the action of the twist flow.

We then consider the interpretation in section \ref{sec:HRTaction}. Applying (\ref{n_a}), (\ref{e_a}), (\ref{delrho}), (\ref{ABCsigma^munu}) and (\ref{ABCK_munu}) on (\ref{twflowid}) yields an interpretation of the action of the twist flow on the initial data
\begin{align}\label{extwid}
&\Delta\sigma_{zz}=o(\lambda)\notag\\
&\Delta\sigma_{zx}=8\pi G\lambda\Big(\frac{1}{z}+\mathcal{O}(z)\Big)\delta(x_0-x)+o(\lambda)\notag\\
&\Delta\sigma_{xx}=o(\lambda)\notag\\
&\Delta K_{ab}=o(\lambda),
\end{align}
where we have used 
\begin{align}
e^mn^nK_{mn}=&\Big[-z(x_0-x)-z^2\sigma_{zx}(x_0-x)^2\Big]K_{zz}+\Big[z(x_0-x)-z^2\sigma_{zx}(x_0-x)^2\Big]K_{xx}\notag\\
&+(x_0-x)^2K_{zx}+\mathcal{O}(z^2)\notag\\
=&\# (x_0-x).
\end{align}

Applying (\ref{extwid}) on (\ref{Dbstid}) yields an interpretation of initial-data form of the action of the twist flow on the boundary stress tensor
\begin{align}\label{twistbst}
	\Delta T^{(0)}_{tt}=&\lim\limits_{\epsilon\to 0}(-2\pi)\Big[\lambda\Big(\delta^\prime(x_0-x)-z\sigma_{zx}\delta(x_0-x)\Big)+o(\lambda)\Big]\Big|_{z=\epsilon}\notag\\
	\Delta T^{(0)}_{tx}=&\lim\limits_{\epsilon\to 0}(-2\pi)\Big(\lambda zK_{xx}\delta(x_0-x)+o(\lambda)\Big)\Big|_{z=\epsilon}\notag\\
	\Delta T^{(0)}_{xx}=&\lim\limits_{\epsilon\to 0}(-2\pi)\Big[\lambda\Big(\delta^\prime(x_0-x)-z\sigma_{zx}\delta(x_0-x)\Big)+o(\lambda)\Big]\Big|_{z=\epsilon},
\end{align}
which, combined with (\ref{delAxt}), further yields the non-trivial components
\be\label{twistbst3} \Delta T_{tt}=\Delta T_{xx}=\lim\limits_{\epsilon\to 0}(-2\pi)\Big(\lambda\delta^\prime(x_0-x)+o(\lambda)\Big)\Big|_{z=\epsilon}, \ee
which are consistent with the results derived in \cite{Wang:2023jfr} with the covariant phase space formalism. 

The algebraic structure of (\ref{actbst}) and (\ref{twistbst}) (or (\ref{actbst3}) and (\ref{twistbst3})) manifests the conjugation between the HRT area and the twist along the geodesic in AdS$_3$.

\subsection{A kink transformation dual to Connes Cocycle flow}
In this section, we revisit the change of the boundary stress tensor under a kink transformation dual to the Connes Cocycle flow in \cite{Bousso:2020yxi} with our approach. 

It was proposed that this kink transformation without boundary-condition-preservation is the gravity dual of connes cocycle flow in \cite{Bousso:2020yxi}. However, it is not direct in \cite{Bousso:2020yxi} to introduce the boundary stress tensor, where they choose a special gauge, Feffermann-Graham gauge, such that some geometric quantities can be viewed as functionals of the boundary stress tensor. Our approach performs a computation on the boundary stress tensor in a straightforward way.

\begin{figure}
  \centering
  \includegraphics[width=16cm]{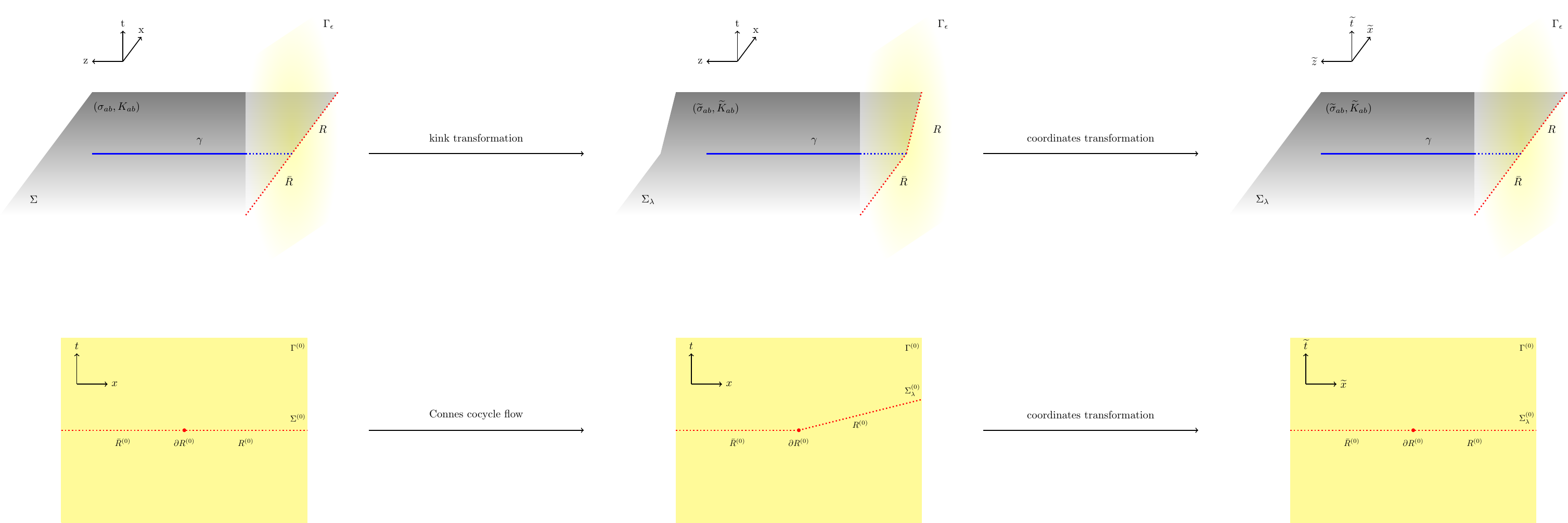}\\
  \caption{A kink transformation dual of connes cocycle flow. The first step shows a kink transformation and its boundary dual, connes cocycle flow. The second step shows a coordinate transformation. Under the new coordinates $\{\widetilde{x}^\mu\}$, the Cauchy surface is located at $\widetilde{t}=0$.}\label{Fig:kink2}
\end{figure}

We review their setup in \cite{Bousso:2020yxi} by considering a kink transformation in AdS$_3$ with flat asymptotic boundary (\ref{flatbdy}) as
\be\label{kink trans} (K_{\Sigma_\lambda})_{ab}=(K_\Sigma)_{ab}+8\pi G\lambda n_an_b\delta(\rho)+o(\lambda), \ee
where we give the initial data on the Cauchy surface $\Sigma$ an evolution such that the evolved initial data $(K_{\Sigma_\lambda})_{ab}$ satisfies (\ref{kink trans}) on a kinked Cauchy surface $\Sigma_\lambda$. Meanwhile, this evolution--a kink transformation--has a boundary dual--connes cocycle flow--which produces a kink transformation on the Cauchy surface $\Sigma^
{(0)}$ in asymptotic boundary, as shown in Fig.\ref{Fig:kink2}. This transformation corresponds to the change of the boundary stress tensor.

We next compute this change of the boundary stress tensor with our approach. For simplicity, we abbreviate $K_\Sigma$ and $K_{\Sigma_\lambda}$ in (\ref{kink trans}) as $K$ and $\widetilde{K}$ in the following, respectively. We denote the boundary stress tensor before and after kink transformation (\ref{kink trans}) by $T_{ab}$ and $\widetilde{T}_{ab}$, respectively.

For the initial system, we still take the interpretation in section \ref{sec:HRTaction} with appropriate coordinates $(z,t,x)$ such that the initial Cauchy surface $\Sigma$ locates at $t=0$ and the HRT surface locates at $x=0$ near the asymptotic boundary. Then we can get the kinked Cauchy surface $\Sigma_\lambda$ located at 
\begin{align}\label{Sigmalamb}
&t=0, \quad\text{one side with}\ x>0,\forall z\notag\\
&t=\Big(-8\pi G\lambda+o(\lambda)\Big)x, \quad\text{the other side with}\ x<0,\forall z.
\end{align}
Note that simultaneously the Cauchy surface $\Sigma^{(0)}$ in the asymptotic boundary transforms into a kinked surface $\Sigma^{(0)}_\lambda$. It allows us to take an appropriate coordinate transformation such that in new coordinate $(\widetilde{z},\widetilde{t},\widetilde{x})$ it looks the same as our previous case in section \ref{sec:BST} and \ref{sec:HRTaction}. More precisely, we can adopt new coordinates $(z,\widetilde{t},\widetilde{x})$ such that the kinked Cauchy surface $\Sigma_\lambda$ (\ref{Sigmalamb}) is located at $\widetilde{t}=0$ and the HRT surface is located at $\widetilde{x}=0$ near the asymptotic boundary. We also adopt a coordinate transformation for the asymptotic boundary coordinates from $(t,x)$ to $(\widetilde{t},\widetilde{x})$.

Combining with (\ref{Sigmalamb}), we explicitly write this coordinate transformation as
\begin{align}
\widetilde{t}+\widetilde{x}=&(t+x)\theta(t+x)+(1+8\pi G\lambda+o(\lambda))(t+x)(1-\theta(t+x))\notag\\
\widetilde{t}-\widetilde{x}=&(1-8\pi G\lambda+o(\lambda))(t-x)\theta(t-x)+(t-x)(1-\theta(t-x)),
\end{align}
and its inverse transformation yields a pullback of asymptotic boundary metric $\eta_{ab}$
\be\label{newg0} \eta_{ab}(t,x)\to \widetilde{g}^{(0)}_{ab}(\widetilde{t}, \widetilde{x})=J(\widetilde{t},\widetilde{x})\eta_{ab}(\widetilde{t},\widetilde{x}), \ee
where the prefactor $J(\widetilde{t},\widetilde{x})$ is
\be\label{J} J(\widetilde{t},\widetilde{x})=1+8\pi G\lambda(\theta(\widetilde{t}-\widetilde{x})+\theta(\widetilde{t}+\widetilde{x})-1)+o(\lambda). \ee

We then can evaluate the boundary stress tensor $\widetilde{T}_{ab}$ in this set of new coordinates. Notice that in this set of new coordinates, we can view the set of Cauchy surfaces $\Sigma_\lambda$ as a foliation of the spacetime similar to the one in section \ref{sec:BST}. Now we mark the quantities in this foliation with a tilde\footnote{Although we also denote the kink transformed initial data $(\widetilde{\sigma}_{ab},\widetilde{K}_{ab})$ with this tilde symbol. We can view them in two aspects: the quantities in this new foliation with new coordinates, the quantities in old foliation with old coordinates. So, if we write $h^{(0)ab}\widetilde{\sigma}_{ab}$, it means that we project the kink transformed initial data $\widetilde{\sigma}_{ab}$ in old foliation with old coordinates.}, e.g., $\widetilde{u}^{(0)}_a$ and $\widetilde{\sigma}_{ab}$. Applying (\ref{newg0}) on the formulas in Appendix \ref{appendix:Foliation} yields some relations on this new foliation
\begin{align}\label{RelaNfolia}
\widetilde{h}^{(0)}_{ab}=&Jh^{(0)}_{ab}\notag\\
\widetilde{u}^{(0)}_{a}=&J^{\frac{1}{2}}u^{(0)}_{a}\notag\\
\widetilde{h}^{(0)ab}=&J^{-1}h^{(0)ab}\notag\\
\widetilde{u}^{(0)a}=&J^{-\frac{1}{2}}u^{(0)a}\notag\\
\widetilde{h}^{(0)b}_{a}=&h^{(0)b}_{a}.
\end{align}
Similar to the result (\ref{Pbstid}) in section \ref{sec:BST}, we write the projection of $\widetilde{T}_{ab}$ in this new foliation as
\begin{align}\label{NPbstid}
\widetilde{u}^{(0)m}\widetilde{u}^{(0)n}\widetilde{T}_{mn}=\lim_{\epsilon\to 0}\frac{1}{4G}\Big[&-\frac{1}{2}(\widetilde{\sigma}_{zz}-\frac{1}{z^2})-(\widetilde{h}^{(0)ab}\widetilde{\sigma}_{ab}-\frac{1}{z^2})-\frac{1}{2}z\partial_z(\widetilde{h}^{(0)ab}\widetilde{\sigma}_{ab}-\frac{1}{z^2})\notag\\
&+z\widetilde{h}^{(0)ab}\widetilde{D}^{(0)}_a(\widetilde{h}^{(0)c}_b\widetilde{\sigma}_{zc})+\frac{1}{2}(\widetilde{K}^{(0)})^2\notag\\
&-\frac{1}{2}z^2(\widetilde{h}^{(0)ab}\widetilde{K}_{ab})^2+\frac{1}{2}z^2\widetilde{h}^{(0)ab}\widetilde{\sigma}_{za}\widetilde{\sigma}_{zb}+\mathcal{O}(z^2)\Big]\Big|_{z=\epsilon}\notag\\
\widetilde{h}^{(0)m}_a\widetilde{u}^{(0)n}\widetilde{T}_{mn}=\lim_{\epsilon\to 0}\frac{1}{4G}\Big[&\widetilde{h}^{(0)b}_a\widetilde{K}_{zb}+\widetilde{h}^{(0)b}_a\widetilde{D}^{(0)}_b(z\widetilde{h}^{(0)cd}\widetilde{K}_{cd}-\widetilde{K}^{(0)})\notag\\
&-z^2\widetilde{h}^{(0)b}_a\widetilde{\sigma}_{zb}\widetilde{h}^{(0)cd}\widetilde{K}_{cd}+\mathcal{O}(z^2)\Big]\Big|_{z=\epsilon}\notag\\
\widetilde{h}^{(0)m}_a\widetilde{h}^{(0)n}_b\widetilde{T}_{mn}=\lim_{\epsilon\to 0}\frac{1}{4G}\Big[&\Big(-\frac{1}{2}(\widetilde{\sigma}_{zz}-\frac{1}{z^2})-(\widetilde{h}^{(0)cd}\widetilde{\sigma}_{cd}-\frac{1}{z^2})-\frac{1}{2}z\partial_z(\widetilde{h}^{(0)cd}\widetilde{\sigma}_{cd}-\frac{1}{z^2})\notag\\
&+z\widetilde{h}^{(0)cd}\widetilde{D}^{(0)}_c(\widetilde{h}^{(0)f}_d\widetilde{\sigma}_{zf})+\frac{1}{2}(\widetilde{K}^{(0)})^2-\frac{1}{2}\widetilde{R}^{(0)}\notag\\
&-\frac{1}{2}z^2(\widetilde{h}^{(0)cd}\widetilde{K}_{cd})^2+\frac{1}{2}z^2\widetilde{h}^{(0)cd}\widetilde{\sigma}_{zc}\widetilde{\sigma}_{zd}\Big)\widetilde{h}^{(0)}_{ab}+\mathcal{O}(z^2)\Big]\Big|_{z=\epsilon}.
\end{align}
And here the projection of $T_{ab}$ (\ref{Pbstid}) in old foliation is the same as the result (\ref{Pbstid}) in section (\ref{sec:BST}) with $g^{(0)}_{ab}=\eta_{ab}$, i.e., 
\begin{align}\label{OPbstid}
u^{(0)m}u^{(0)n}T_{mn}=\lim_{\epsilon\to 0}\frac{1}{4G}\Big[&-\frac{1}{2}(\sigma_{zz}-\frac{1}{z^2})-(h^{(0)ab}\sigma_{ab}-\frac{1}{z^2})-\frac{1}{2}z\partial_z(h^{(0)ab}\sigma_{ab}-\frac{1}{z^2})\notag\\
&+zh^{(0)ab}D^{(0)}_a(h^{(0)c}_b\sigma_{zc})-\frac{1}{2}z^2(h^{(0)ab}K_{ab})^2\notag\\
&+\frac{1}{2}z^2h^{(0)ab}\sigma_{za}\sigma_{zb}+\mathcal{O}(z^2)\Big]\Big|_{z=\epsilon}\notag\\
h^{(0)m}_au^{(0)n}T_{mn}=\lim_{\epsilon\to 0}\frac{1}{4G}\Big[&h^{(0)b}_a K_{zb}+zh^{(0)b}_aD^{(0)}_b(h^{(0)cd}K_{cd})-z^2h^{(0)b}_a\sigma_{zb}h^{(0)cd}K_{cd}+\mathcal{O}(z^2)\Big]\Big|_{z=\epsilon}\notag\\
h^{(0)m}_ah^{(0)n}_bT_{mn}=\lim_{\epsilon\to 0}\frac{1}{4G}\Big[&\Big(-\frac{1}{2}(\sigma_{zz}-\frac{1}{z^2})-(h^{(0)cd}\sigma_{cd}-\frac{1}{z^2})-\frac{1}{2}z\partial_z(h^{(0)cd}\sigma_{cd}-\frac{1}{z^2})\notag\\
&+zh^{(0)cd}D^{(0)}_c(h^{(0)f}_d\sigma_{zf})-\frac{1}{2}z^2(h^{(0)cd}K_{cd})^2\notag\\
&+\frac{1}{2}z^2h^{(0)cd}\sigma_{zc}\sigma_{zd}\Big)h^{(0)}_{ab}+\mathcal{O}(z^2)\Big]\Big|_{z=\epsilon}.
\end{align}
We then present the change of the boundary stress tensor as
\begin{align}\label{changebst}
&\widetilde{u}^{(0)m}\widetilde{u}^{(0)n}\widetilde{T}_{mn}-u^{(0)m}u^{(0)n}T_{mn}\notag\\
&\widetilde{h}^{(0)m}_a\widetilde{u}^{(0)n}\widetilde{T}_{mn}-h^{(0)m}_au^{(0)n}T_{mn}\notag\\
&\widetilde{h}^{(0)m}_a\widetilde{h}^{(0)n}_b\widetilde{T}_{mn}-h^{(0)m}_ah^{(0)n}_bT_{mn}.
\end{align}
Applying (\ref{kink trans}), (\ref{J}), (\ref{RelaNfolia}), (\ref{NPbstid}) and (\ref{OPbstid}) on (\ref{changebst}) and adopting $u^{(0)}_t=\widetilde{u}^{(0)}_{\widetilde{t}}=1,u^{(0)}_x=\widetilde{u}^{(0)}_{\widetilde{x}}=0$ yields an explicit form of the change of the boundary stress tensor\footnote{We also have used that $\delta(x)=\delta(\widetilde{x})$ since the HRT surface locates on $x=0$ and $\widetilde{x}=0$ in old and new coordinates, respectively.}
\begin{align}\label{changebst2}
\widetilde{T}_{\tilde{t}\tilde{t}}-T_{tt}=&\lim_{\epsilon\to 0}\Big\{-2\pi\lambda zK_{xx}\delta(x)+o(\delta,\lambda)\Big\}\Big|_{z=\epsilon}\notag\\
\widetilde{T}_{\tilde{t}\tilde{x}}-T_{tx}=&\lim_{\epsilon\to 0}\Big\{-2\pi\lambda z\sigma_{zx}\delta(x)+o(\delta,\lambda)\Big\}\Big|_{z=\epsilon}\notag\\
\widetilde{T}_{\tilde{x}\tilde{x}}-T_{xx}=&\lim_{\epsilon\to 0}\Big\{-2\pi\lambda zK_{xx}\delta(x)+o(\delta,\lambda)\Big\}\Big|_{z=\epsilon},
\end{align}
where $o(\delta,\lambda)$ contains the finite terms on the HRT surface $x=0$ and higher order terms of $\lambda$, and in using (\ref{kink trans}), we actually use the following relations
\begin{align}
\widetilde{\sigma}_{zz}=&\sigma_{zz}\notag\\
h^{(0)b}_a\widetilde{\sigma}_{zb}=&h^{(0)b}_a\sigma_{zb}\notag\\
h^{(0)ab}\widetilde{\sigma}_{ab}=&h^{(0)ab}\sigma_{ab}
\end{align}
and
\begin{align}
\widetilde{K}_{zz}=&K_{zz}\notag\\
h^{(0)b}_a\widetilde{K}_{zb}=&h^{(0)b}_aK_{za}\notag\\
h^{(0)c}_ah^{(0)d}_b\widetilde{K}_{cd}=&h^{(0)c}_ah^{(0)d}_bK_{ab}+8\pi G\lambda z^{-1}\delta(x)h^{(0)c}_ah^{(0)d}_bK_{cd}.
\end{align}
Applying (\ref{delAxt}) and (\ref{EE}) on (\ref{changebst2}), we represent the change of the boundary stress tensor in terms of the entanglement entropy
\be\label{change of bst 3} \widetilde{T}_{\tilde{t}\tilde{x}}-T_{tx}=-8\pi G\lambda \frac{\delta S}{\delta x}\delta(x)+o(\delta,\lambda) \ee
which is consistent with the result in \cite{Bousso:2020yxi}.


\section*{Acknowledgments}

We thank for the discussion with Gabriel Arenas-Henriquez, Liangyu Chen, Yiming Chen, Hong Liu, Baishali Roy, Yu-Xuan Zhang. X.s.W thanks the 19th Asian Winter School on Strings, Particles and Cosmology. X.s.W thanks the workshop on "Black Hole, Quantum Chaos and Quantum Information" held at the Yukawa Institute for Theoretical Physics at Kyoto University. X.s.W thanks the China-India-UK School in Mathematical Physics, held by ICMS, Edinburgh.

X.s.W. and J.q.W. are supported by the National Natural Science Foundation of China (NSFC) Project No.12447101.

\appendix

\section{Foliation}\label{appendix:Foliation}
In this appendix, we introduce the twin foliation of the bulk spacetime and the asymptotic boundary, as shown in Fig.\ref{Fig:foliation}. 

For convenience, we first foliate the asymptotic boundary $\Gamma^{(0)}$ with a set of Cauchy surfaces $\Sigma^{(0)}_{T}$ labelled by a scalar field $T^{(0)}$. Based on the scalar field $T^{(0)}$, we introduce a unit future-pointing normal vector $\hat{u}^{(0)}$ to the Cauchy surface $\Sigma^{(0)}_{T}$ by
\be\label{u_a} u^{(0)}_a=-\frac{\partial_a T^{(0)}}{\sqrt{-g^{(0)mn}\partial_mT^{(0)}\partial_nT^{(0)}}}, \ee
and
\be\label{u^a} u^{(0)a}=g^{(0)ab}u^{(0)}_b, \ee
with
\be\label{unitu} g^{(0)}_{ab}u^{(0)a}u^{(0)b}=-1. \ee
It also serves as a bridge between the metric $g^{(0)}_{ab}$ of the asymptotic boundary $\Gamma^{(0)}$ and the induced metric $h^{(0)}_{ab}$ on the Cauchy surface $\Sigma^{(0)}_{T}$
\be\label{hab} h^{(0)}_{ab}=g^{(0)}_{ab}+u^{(0)}_au^{(0)}_b. \ee
We introduce the acceleration of $\Sigma^{(0)}_T$ on $\Gamma^{(0)}$ by
\be\label{a0} a^{(0)}_b=u^{(0)c}D^{(0)}_cu^{(0)}_b. \ee
We introduce the extrinsic curvature $K^{(0)}_{ab}$ of the Cauchy surface $\Sigma^{(0)}_{T}$ in the asymptotic boundary $\Gamma^{(0)}$ by
\be\label{K0ab} K^{(0)}_{ab}=h^{(0)c}_aD^{(0)}_cu^{(0)}_b, \ee
where $D^{(0)}$ denotes the covariant derivative associated with the metric $g^{(0)}_{ab}$, and its trace as
\be\label{K0} K^{(0)}=h^{(0)ab} K^{(0)}_{ab}. \ee
 
We then foliate the bulk spacetime with a set of Cauchy surfaces $\Sigma_{T}$ labelled by a scalar field $T$, where we mainly focus on the region near the asymptotic boundary since this region contains the essential physics in AdS$_3$. Here  we assume that in the near asymptotic boundary region with ($z,x^a$) coordinates, the scalar field $T$ is independent of $z$. Then it is natural to set\footnote{Here we only do this matching before the action of the HRT area flow. We should emphasize that the HRT area flow does not act on the $\Sigma^{(0)}_{T}$ since its action is a boundary-condition-preserving action.}
\be\label{match} T=T^{(0)}, \ee
to match the two foliations. Parallel to the foliation of the asymptotic boundary, we introduce a unit future-pointing normal vector $\hat{\tau}$ by
\be\label{tau_mu} \tau_\mu=-\frac{\partial_\mu T^{(0)}}{\sqrt{-g^{(0)\alpha\beta}\partial_\alpha T^{(0)}\partial_\beta T^{(0)}}}, \ee
and 
\be\label{tau^mu} \tau^\mu=g^{\mu\nu}\tau_\nu, \ee
where we have used the matching condition (\ref{match}). We then introduce the induced metric $\sigma_{\mu\nu}$ on the Cauchy surface $\Sigma_{T}$ by
\be\label{sigma_munu} \sigma_{\mu\nu}=g_{\mu\nu}+\tau_\mu\tau_\nu. \ee
We introduce the extrinsic curvature $K_{\mu\nu}$ of the Cauchy surface $\Sigma_{T}$ in the bulk spacetime by
\be\label{barK_munu} K_{\mu\nu}=\sigma^{\rho}_\mu\nabla_\rho\tau_\nu, \ee
where $\nabla$ denotes the covariant derivative associated with the metric $g_{ab}$, and its trace as
\be\label{barK} K=\sigma^{\mu\nu}K_{\mu\nu}. \ee

\section{The asymptotic behaviors}\label{appendix:asy}
In this appendix, we compute the asymptotic behaviors of some quantities from (\ref{ABC0}). 

We first list the asymptotic form of some quantities directly related to the metric (\ref{ABC0}). 

The inverse metric components $g^{\mu\nu}$ are
\begin{align}\label{ABCig}
g^{zz}=&z^2-z^4(g_{zz}-\frac{1}{z^2})+z^6g^{(0)ab}g_{za}g_{zb}+\mathcal{O}(z^6)\notag\\
g^{za}=&-z^4g^{(0)ab}g_{zb}+z^6g^{(0)ab}g_{zb}(g_{zz}-\frac{1}{z^2})\notag\\
&+z^6g^{(0)ac}g^{(0)bd}(g_{cd}-\frac{1}{z^2}g^{(0)}_{cd})g_{zb}-z^8g^{(0)ac}g^{(0)bd}g_{zc}g_{zd}g_{zb}+\mathcal{O}(z^7)\notag\\
g^{ab}=&z^2g^{(0)ab}-z^4g^{(0)ac}g^{(0)bd}(g_{cd}-\frac{1}{z^2}g^{(0)}_{cd})+z^6g^{(0)ac}g^{(0)bd}g_{zc}g_{zd}+\mathcal{O}(z^6),
\end{align}
where $g^{(0)ab}$ denotes the inverse of $g^{(0)}_{ab}$.

The Christoffel connection components $\Gamma^\rho_{\mu\nu}$ are
\begin{align}\label{ABCGamma}
\Gamma^z_{zz}=&-\frac{1}{z}+z(g_{zz}-\frac{1}{z^2})+\frac{1}{2}z^2\partial_z(g_{zz}-\frac{1}{z^2})-z^3g^{(0)ab}g_{za}g_{zb}-z^4g^{(0)ab}g_{za}\partial_zg_{zb}+\mathcal{O}(z^3)\notag\\
\Gamma^z_{za}=&zg_{za}+\frac{1}{2}z^2\partial_a(g_{zz}-\frac{1}{z^2})-z^3(g_{zz}-\frac{1}{z^2})g_{za}\notag\\
&-z^3g^{(0)bc}g_{zc}(g_{ab}-\frac{1}{z^2}g^{(0)}_{ab})-\frac{1}{2}z^4g^{(0)bc}g_{zc}\partial_z(g_{ab}-\frac{1}{z^2}g^{(0)}_{ab})\notag\\
&-\frac{1}{2}z^4g^{(0)bc}g_{zc}(D^{(0)}_ag_{zb}-D^{(0)}_bg_{za})+z^5g^{(0)bc}g_{zb}g_{zc}g_{za}+\mathcal{O}(z^4)\notag\\
\Gamma^z_{ab}=&\frac{1}{z}g^{(0)}_{ab}-z(g_{zz}-\frac{1}{z^2})g^{(0)}_{ab}-\frac{1}{2}z^2\partial_z(g_{ab}-\frac{1}{z^2}g^{(0)}_{ab})\notag\\
&+\frac{1}{2}z^2(D^{(0)}_ag_{zb}+D^{(0)}_bg_{za})+z^3g^{(0)cd}g_{zc}g_{zd}g^{(0)}_{ab}+\mathcal{O}(z^3)\notag\\
\Gamma^a_{zz}=&zg^{(0)ab}g_{zb}+z^2g^{(0)ab}\partial_zg_{zb}\notag\\
&-\frac{1}{2}z^2g^{(0)ab}\partial_b(g_{zz}-\frac{1}{z^2})-z^3g^{(0)ab}g_{zb}(g_{zz}-\frac{1}{z^2})-\frac{1}{2}z^4g^{(0)ab}g_{zb}\partial_z(g_{zz}-\frac{1}{z^2})\notag\\
&-z^3g^{(0)ac}g^{(0)bd}(g_{cd}-\frac{1}{z^2}g^{(0)}_{cd})g_{zb}-z^4g^{(0)ac}g^{(0)bd}(g_{cd}-\frac{1}{z^2}g^{(0)}_{cd})\partial_zg_{zb}\notag\\
&+z^5g^{(0)ac}g^{(0)bd}g_{zc}g_{zd}g_{zb}+z^6g^{(0)ac}g^{(0)bd}g_{zc}g_{zd}\partial_zg_{zb}+\mathcal{O}(z^4)\notag\\
\Gamma^a_{zb}=&-\frac{1}{z}\delta^a_b+zg^{(0)ac}(g_{bc}-\frac{1}{z^2}g^{(0)}_{bc})+\frac{1}{2}z^2g^{(0)ac}\partial_z(g_{bc}-\frac{1}{z^2}g^{(0)}_{bc})\notag\\
&+\frac{1}{2}z^2g^{(0)ac}(D^{(0)}_bg_{zc}-D^{(0)}_cg_{zb})-z^3g^{(0)ac}g_{zc}g_{zb}+\mathcal{O}(z^3)\notag\\
\Gamma^a_{bc}=&\Gamma^{(0)a}_{bc}-zg^{(0)ad}g_{zd}g^{(0)}_{bc}\notag\\
&+\frac{1}{2}z^2g^{(0)ad}\Big[D^{(0)}_c(g_{bd}-\frac{1}{z^2}g^{(0)}_{bd})+D^{(0)}_b(g_{cd}-\frac{1}{z^2}g^{(0)}_{cd})-D^{(0)}_d(g_{bc}-\frac{1}{z^2}g^{(0)}_{bc})\Big]\notag\\
&+z^3g^{(0)ad}g_{zd}(g_{zz}-\frac{1}{z^2})g^{(0)}_{bc}+z^3g^{(0)ad}g^{(0)fh}(g_{dh}-\frac{1}{z^2}g^{(0)}_{dh})g_{zf}g^{(0)}_{bc}\notag\\
&-\frac{1}{2}z^4g^{(0)ad}g_{zd}(D^{(0)}_cg_{zb}+D^{(0)}_bg_{zc})+\frac{1}{2}z^4g^{(0)ad}g_{zd}\partial_z(g_{bc}-\frac{1}{z^2}g^{(0)}_{bc})\notag\\
&-z^5g^{(0)ad}g^{(0)hf}g_{zd}g_{zh}g_{zf}g^{(0)}_{bc}+\mathcal{O}(z^4),
\end{align}
where $\Gamma^{(0)c}_{ab}$ denotes the connection associated with $g^{(0)}_{ab}$
\begin{align}
\Gamma^{(0)c}_{ab}=\frac{1}{2}g^{(0)cd}(\partial_bg^{(0)}_{ad}+\partial_ag^{(0)}_{bd}-\partial_dg^{(0)}_{ab}).
\end{align}

We then list the asymptotic form of some quantities with holographic renormalization.

The induced metric on $\Gamma_z$ are
\begin{align}\label{ABCgamma}
	\gamma_{zz}=&z^2g^{(0)ab}g_{za}g_{zb}+\mathcal{O}(z^2)\notag\\
	\gamma_{za}=&g_{za}\notag\\
	\gamma_{ab}=&g_{ab}.
\end{align}

The extrinsic curvature of $\Gamma_z$ are
\begin{align}\label{ABCKGamma}
	\mathcal{K}_{zz}=&z^2g^{(0)ab}g_{za}g_{zb}+\mathcal{O}(z^2)\notag\\
	\mathcal{K}_{za}=&g_{za}+\mathcal{O}(z)\notag\\
	\mathcal{K}_{ab}=&\frac{1}{z^2}g^{(0)}_{ab}-\frac{1}{2}(g_{zz}-\frac{1}{z^2})g^{(0)}_{ab}+\frac{1}{2}z^2g^{(0)cd}g_{zc}g_{zd}g^{(0)}_{ab}\notag\\
	&+\frac{1}{2}z(D^{(0)}_ag_{zb}+D^{(0)}_bg_{za})-\frac{1}{2}z\partial_z(g_{ab}-\frac{1}{z^2}g^{(0)}_{ab})+\mathcal{O}(z^2).
\end{align}

The extrinsic curvature scalar of $\Gamma_z$ is
\begin{align}\label{ABCKG}
	\mathcal{K}=&\gamma^{ab}\mathcal{K}_{ab}\notag\\
	=&2-z^2(g_{zz}-\frac{1}{z^2})-z^2g^{(0)ab}(g_{ab}-\frac{1}{z^2}g^{(0)}_{ab})+z^4g^{(0)ab}g_{za}g_{zb}\notag\\
	&+z^3g^{(0)ab}D^{(0)}_ag_{zb}-\frac{1}{2}z^3g^{(0)ab}\partial_z(g_{ab}-\frac{1}{z^2}g^{(0)}_{ab})+\mathcal{O}(z^4).
\end{align}

By virtue of the twin foliation in Appendix \ref{appendix:Foliation}, we can further compute the asymptotic form of some quantities appearing in the foliation of bulk spacetime and represent them in terms of some quantities appearing in the foliation of the asymptotic boundary. 

More precisely, applying (\ref{ABC0}) on (\ref{tau_mu}), combined with (\ref{u_a}), yields the asymptotic form of $\tau_\mu$
\begin{align}\label{ABCtau_mu}
\tau_z=&0\notag\\
\tau_a=&\frac{1}{z}u^{(0)}_a-\frac{1}{2}zu^{(0)}_au^{(0)b}u^{(0)c}(g_{bc}-\frac{1}{z^2}g^{(0)}_{bc})+\frac{1}{2}z^3u^{(0)}_a(u^{(0)b}g_{zb})^2+\mathcal{O}(z^3),
\end{align}
which further, combined with (\ref{tau^mu}) and (\ref{ABCig}), yields the asymptotic form of $\tau^\mu$
\begin{align}\label{ABCtau^mu}
\tau^z=&-z^3u^{(0)a}g_{za}+z^5u^{(0)a}g_{za}(g_{zz}-\frac{1}{z^2})+z^5u^{(0)a}h^{(0)bc}(g_{ab}-\frac{1}{z^2}g^{(0)}_{ab})g_{zc}\notag\\
&-\frac{1}{2}z^5u^{(0)a}u^{(0)b}u^{(0)c}(g_{ab}-\frac{1}{z^2}g^{(0)}_{ab})g_{zc}-z^7u^{(0)a}h^{(0)bc}g_{za}g_{zb}g_{zc}+\frac{1}{2}z^7(u^{(0)a}g_{za})^3+\mathcal{O}(z^6)\notag\\
\tau^a=&zu^{(0)a}-z^3h^{(0)ab}u^{(0)c}(g_{bc}-\frac{1}{z^2}g^{(0)}_{bc})+\frac{1}{2}z^3u^{(0)a}u^{(0)b}u^{(0)c}(g_{bc}-\frac{1}{z^2}g^{(0)}_{bc})\notag\\
&+z^5h^{(0)ab}u^{(0)c}g_{zb}g_{zc}-\frac{1}{2}z^5u^{(0)a}(u^{(0)b}g_{zb})^2+\mathcal{O}(z^5).
\end{align}

Applying (\ref{ABC0}) and (\ref{ABCtau_mu}) on (\ref{sigma_munu}) yields the asymptotic form of $\sigma_{\mu\nu}, \sigma^\mu_\nu, \sigma^{\mu\nu}$ as
\begin{align}\label{ABCsigma_munu}
\sigma_{zz}=&\frac{1}{z^2}+(g_{zz}-\frac{1}{z^2})\notag\\
\sigma_{za}=&h^{(0)b}_ag_{zb}-u^{(0)}_au^{(0)b}g_{zb}\notag\\
\sigma_{ab}=&\frac{1}{z^2}h^{(0)}_{ab}+h^{(0)}_{ab}h^{(0)cd}(g_{cd}-\frac{1}{z^2}g^{(0)}_{cd})-u^{(0)}_ah^{(0)c}_bu^{(0)d}(g_{cd}-\frac{1}{z^2}g^{(0)}_{cd})\notag\\
&-u^{(0)}_bh^{(0)c}_au^{(0)d}(g_{cd}-\frac{1}{z^2}g^{(0)}_{cd})+z^2u^{(0)}_au^{(0)}_b(u^{(0)c}g_{zc})^2+\mathcal{O}(z^2),
\end{align}
and
\begin{align}\label{ABCsigma_mu^nu}
\sigma^z_z=&1\notag\\
\sigma^a_z=&0\notag\\
\sigma^z_a=&-z^2u^{(0)}_au^{(0)b}g_{zb}+z^4u^{(0)}_au^{(0)b}g_{zb}(g_{zz}-\frac{1}{z^2})\notag\\
&+z^4u^{(0)}_au^{(0)b}h^{(0)cd}(g_{bc}-\frac{1}{z^2}g^{(0)}_{bc})g_{zd}-z^6u^{(0)}_au^{(0)b}h^{(0)cd}g_{zb}g_{zc}g_{zd}+\mathcal{O}(z^5)\notag\\
\sigma^b_a=&h^{(0)b}_a-z^2u^{(0)}_ah^{(0)bc}u^{(0)d}(g_{cd}-\frac{1}{z^2}g^{(0)}_{cd})+z^4u^{(0)}_ah^{(0)bc}u^{(0)d}g_{zc}g_{zd}+\mathcal{O}(z^4),
\end{align}
and
\begin{align}\label{ABCsigma^munu}
\sigma^{zz}=&z^2-z^4(g_{zz}-\frac{1}{z^2})+z^6h^{(0)ab}g_{za}g_{zb}+\mathcal{O}(z^6)\notag\\
\sigma^{za}=&-z^4h^{(0)ab}g_{zb}+z^6h^{(0)ab}g_{zb}(g_{zz}-\frac{1}{z^2})+z^6h^{(0)ab}h^{(0)cd}(g_{cd}-\frac{1}{z^2}g^{(0)}_{cd})g_{zb}\notag\\
&-z^8h^{(0)ab}h^{(0)cd}g_{zb}g_{zc}g_{zd}+\mathcal{O}(z^7)\notag\\
\sigma^{ab}=&z^2h^{(0)ab}-z^4h^{(0)ab}h^{(0)cd}(g_{cd}-\frac{1}{z^2}g^{(0)}_{cd})+z^6h^{(0)ac}h^{(0)bd}g_{zc}g_{zd}+\mathcal{O}(z^6),
\end{align}
where we have used an identity\footnote{One can easily check this identity by acting $h^{(0)ab}$ on the both side of the identity, and noticing that $h^{(0)ab}$ can be viewed as the induced metric on one dimensional manifold.}
\be\label{dim1} h^{(0)m}_ah^{(0)n}_bS_{mn}=h^{(0)}_{ab}h^{(0)mn}S_{mn}, \ee
for a given tensor $S_{mn}$. 

Applying (\ref{ABCGamma}), (\ref{ABCtau_mu}) and (\ref{ABCsigma_mu^nu}) on (\ref{barK_munu}) yields the asymptotic form of $K_{\mu\nu}$ as
\begin{align}\label{ABCK_munu}
K_{zz}=&-u^{(0)a}g_{za}-zu^{(0)a}\partial_zg_{za}+\mathcal{O}(z)\notag\\
K_{za}=&-h^{(0)b}_au^{(0)c}(g_{bc}-\frac{1}{z^2}g^{(0)}_{bc})-\frac{1}{2}zh^{(0)b}_au^{(0)c}\partial_z(g_{bc}-\frac{1}{z^2}g^{(0)}_{bc})\notag\\
&+\frac{1}{2}zK^{(0)}h^{(0)b}_ag_{zb}-\frac{1}{2}zh^{(0)b}_aD^{(0)}_b(u^{(0)c}g_{zc})+\frac{1}{2}zh^{(0)b}_au^{(0)c}D^{(0)}_c(h^{(0)d}_bg_{zd})\notag\\
&-\frac{1}{2}za^{(0)}_au^{(0)b}g_{zb}+z^2u^{(0)}_a(u^{(0)b}g_{zb})^2+z^2h^{(0)b}_ag_{zb}u^{(0)c}g_{zc}+z^3u^{(0)}_au^{(0)b}u^{(0)c}g_{zb}\partial_zg_{zc}+\mathcal{O}(z^2)\notag\\
K_{ab}=&\frac{1}{z}K^{(0)}_{ab}+h^{(0)}_{ab}u^{(0)c}g_{zc}+\mathcal{O}(z).
\end{align}
Applying (\ref{ABCK_munu}) and (\ref{ABCsigma^munu}) on (\ref{barK}), combined with (\ref{K0}), further yields the asymptotic form of the extrinsic curvature scalar as
\be\label{ABCK} K=zK^{(0)}-z^3u^{(0)a}\partial_zg_{za}+\mathcal{O}(z^3). \ee

\section{A clarification}\label{appendix:clarify}

In this appendix, we clarify how we derive (\ref{actbst3}) from (\ref{actbst}). 

Here, we attempt to reveal a more direct resonance between the HRT area and the initial data appearing in (\ref{actbst}). We first consider the variation of the length of a geodesic with two endpoints on the spatial boundary with respect to the boundary coordinates\footnote{Keep in mind that the main contribution to the variation of the length of the geodesic is from the regions near the endpoints. So it is consistent with our interpretation in section \ref{sec:HRTaction}.}
\be\label{delA} \delta A_{\text{HRT}}=\delta\int_{s_i}^{s_f} ds\sqrt{g_{\mu\nu}\frac{dx^\mu}{ds}\frac{dx^\nu}{ds}}=g_{\mu\nu}e^\mu\delta x^\nu\Big|_{i}^{f}, \ee
where we denote the intersections of the geodesic with IR cutoff surface $z=\epsilon$ by $i$ and $f$, respectively. And we denote the corresponding points at the asymptotic boundary of $i$ and $f$ by $b_i$ and $b_f$, respectively, where we further assume that
\be\label{asyx} \delta x^\mu|_{b_{i(f)}}=\delta x^\mu|_{i(f)}+\mathcal{O}(\epsilon^2). \ee 

In addition, the parametric equation of the geodesic (\ref{line}) and the normalization of $\hat{e}$ tell us the asymptotic behavior of the tangent vector $\hat{e}$ to the geodesic
\be\label{ABCe^mu} e^z=z+\mathcal{O}(z^3),\quad e^t=e^x=0, \ee
which is consistent with the one that we derive from restricting (\ref{e_a}) on the geodesic, raising its index and pushforwarding it to the spacetime.

Applying (\ref{ABC0}), (\ref{asyx}) and (\ref{ABCe^mu}) on (\ref{delA}), we can find that the variation of the length of the geodesic is related to the initial data by 
\begin{align}\label{delAxt}
	0=\frac{\delta A_{\text{HRT}}}{\delta x(s)}\Big|_{b_{i(f)}}=&\lim\limits_{\epsilon\to 0}\frac{\delta A_{\text{HRT}}}{\delta x(s)}\Big|_{{i(f)}}=\lim\limits_{\epsilon\to 0}(g_{zx}e^z)\Big|_{z=\epsilon}=\lim\limits_{\epsilon\to 0}(z\sigma_{zx})\Big|_{z=\epsilon}\notag\\
	0=\frac{\delta A_{\text{HRT}}}{\delta t(s)}\Big|_{b_{i(f)}}=&\lim\limits_{\epsilon\to 0}\frac{\delta A_{\text{HRT}}}{\delta t(s)}\Big|_{{i(f)}}=\lim\limits_{\epsilon\to 0}(g_{zt}e^z)\Big|_{z=\epsilon}=\lim\limits_{\epsilon\to 0}(zK_{xx})\Big|_{z=\epsilon},
\end{align}
where the first equations in both lines arise from the fact that we are considering the boundary region $R$ to be a half line $x\in[x_0,\infty)$ at $t=t_0$, which leads to the length of the geodesic independent of the specific position of the endpoints.

Applying (\ref{delAxt}) on (\ref{actbst}) yields the only non-trivial components 
\be \Delta T^{(0)}_{tx}=\lim\limits_{\epsilon\to 0}(-2\pi)\Big(\lambda\delta^\prime(x_0-x)+o(\lambda)\Big)\Big|_{z=\epsilon}. \ee

\end{document}